\documentclass[12pt]{article}
\usepackage{epsfig}
\usepackage{amsfonts}
\usepackage{amsopn}
\usepackage{amsmath}
\usepackage{verbatim,amsthm}

\title{Construction of the Zero-Energy State of $SU(2)$-Matrix Theory: Near the Origin}

\author{
Jens Hoppe$^a$ \thanks{e-mail:  hoppe@kth.se} \ ,  \
Douglas Lundholm$^a$  \thanks{e-mail: dogge@math.kth.se} \\
and Maciej Trzetrzelewski$^{a,b}$ \thanks{e-mail: 33lewski@th.if.uj.edu.pl}  \\
\\
$^a$  Department of Mathematics,\\
Royal Institute of Technology, \\
KTH, 100 44 Stockholm, \\
Sweden \\ \\
$^b$ Institute of Physics,\\
Jagiellonian University, \\
Reymonta 4, 30-059 Krak\'ow,\\
Poland
}

\date{}

\begin{document}

\maketitle

\begin{abstract}
We explicitly construct a (unique) $Spin(9) \times SU(2)$ singlet
state, $\phi$, involving only the fermionic degrees of freedom of
the supersymmetric matrix-model corresponding to reduced
10-dimensional super Yang-Mills theory, resp. supermembranes in
11-dimensional Minkowski space. Any non-singular wavefunction
annihilated by the 16 supercharges of $SU(2)$ matrix theory must, at
the origin (where it is assumed to be non-vanishing) reduce to
$\phi$.
\end{abstract}

\section{Introduction}

The fermionic degrees of freedom of $SU(2)$-matrix theory (see e.g.
\cite{BFSS}) are three $Spin(9)$ spinors, $\theta_{\hat{\alpha}A}$,
($\hat{\alpha}=1,\ldots,16$, $A=1,2,3$,
$\theta_{\hat{\alpha}A}^{\dagger}=\theta_{\hat{\alpha}A}$),
satisfying canonical anti-commutation relations
\[
\{ \theta_{\hat{\alpha}A},\theta_{\hat{\beta}B} \}=\delta_{\hat{\alpha}\hat{\beta}}\delta_{AB}.
\]
The corresponding $2^{8 \cdot 3}$- dimensional Hilbert-space
$\mathcal{H}=\mathcal{H}_{256} \otimes \mathcal{H}_{256} \otimes
\mathcal{H}_{256}$ splits into irreducible $Spin(9)$ representations built out
of the ones occurring in
\[
\mathcal{H}_{256}=\bold{44}\oplus \bold{84} \oplus \bold{128}. \label{split}
\]
First determining  all
$Spin(9)$ singlets occurring in $\mathcal{H}$, in terms of the three
representations $\bold{44}$, $\bold{84}$, $\bold{128}$ (whose elements are denoted
$|st \rangle$, $|stu \rangle$ and $|t\hat{\alpha} \rangle$ respectively), the central part of the paper then is the explicit construction,
out of these $Spin(9)$ singlets, of a (unique) $Spin(9)\times SU(2)$ singlet $\phi$ (whose
relevance has been advocated by Wosiek, who was led to the
existence of such a  $Spin(9)\times SU(2)$  singlet using symbolic
programme \cite{Wosiek}).

In the next section we take an independent route to obtain $\phi$,
here proving its uniqueness, by listing all possible $Spin(7)\times
SU(2)$ invariant states and then taking
their (unique) linear combination such that the result is
$Spin(9)\times SU(2)$ invariant.

While the three representations in $\mathcal{H}_{256}$, forming an 'Euler-triple' (cp. e.g. \cite{Ramond})
- quite likely relevant concerning the existence of a (unique) zero
energy state for general $SU(N \ge 2)$ (note the intertwining nature
of the two terms
$\gamma^t_{\hat{\beta}\hat{\alpha}}\theta_{\hat{\alpha} A}$ and
$\gamma^{st}_{\hat{\beta}\hat{\alpha}}\theta_{\hat{\alpha} A}$ in
the supercharges of the model) - have quite a long history in supergravity
theory (starting with \cite{CJS}),  we could not \footnote{at least
not until September 28; our apologies to the authors of \cite{PW},
\cite{MTV} (whose results seem to be related to our Fock space representations given in Appendix A)} find a good reference for
their Fock space representations and therefore derived them
explicitly (see Appendix A) to be sure of the exact intertwining
relations.

\section{The construction of $\phi$ }

According to the decomposition of $\mathcal{H}_{256} \otimes
\mathcal{H}_{256} $ into irreducible representations of $Spin(9)$
(cp. e.g. \cite{BHS}, eq. (13)-(18), or \cite{Slansky}) - yielding three $\bold{44}$'s and three
$\bold{84}$'s, which are easily seen to be (proportional to)
\[
\mathop{||}_{44} st \rangle := |su\rangle |tu \rangle  + |tu\rangle |su \rangle - \frac{2}{9} \delta_{st}|uv\rangle |uv \rangle,
\]
\[
\mathop{||}_{84} st \rangle := |suv\rangle |tuv \rangle  + |tuv\rangle |suv \rangle - \frac{2}{9} \delta_{st}|uvw\rangle |uvw \rangle,
\]
\[
\mathop{||}_{128} st \rangle := |s\hat{\alpha}\rangle |t\hat{\alpha}
\rangle  + |t\hat{\alpha} \rangle |s\hat{\alpha} \rangle -
\frac{2}{9} \delta_{st}|u\hat{\alpha} \rangle |u\hat{\alpha}
\rangle,   \label{3.1}
\]
and (for notational convenience we will now write $\alpha$ instead
of $\hat{\alpha}$, in this section)
\[
\mathop{||}_{84} stu \rangle := \epsilon^{stupqrabc}|pqr \rangle |abc \rangle,
\]
\[
\mathop{||}_{128} stu \rangle :=  \gamma^s_{\alpha \beta}|t\alpha \rangle
|u\beta \rangle  +   \gamma^t_{\alpha \beta}|u\alpha \rangle |s\beta
\rangle +  \gamma^u_{\alpha \beta}|s\alpha \rangle |t\beta \rangle,
\]
\[
\mathop{||}_{128} stu \rangle' := \gamma^{stu}_{\alpha \beta}| v \alpha \rangle | v \beta \rangle,  \label{3.2}
\]
there are 14 $Spin(9)$ singlets in $\mathcal{H}_{256} \otimes
\mathcal{H}_{256} \otimes \mathcal{H}_{256}$. Nine of these involve
the $\bold{128}$-dimensional spinor-representations while the
simplest ones are
\[
\mathop{|||}_{44}1\rangle :=|su \rangle_1 |tu \rangle_2 |st \rangle_3, \ \ \ \
\mathop{|||}_{84}1\rangle := \epsilon^{stupqrabc}|stu \rangle_1 |pqr \rangle_2 |abc \rangle_3 \label{3.3},
\]
and the (cyclically invariant) sum of the remaining three,
\[
\mathop{|||}_{844}1\rangle := |suv \rangle_1 |tuv \rangle_2 |st \rangle_3 +   |tuv \rangle_1 |st \rangle_2 |suv \rangle_3  +   |st \rangle_1 |suv \rangle_2 |tuv \rangle_3 . \label{3.4}
\]
Using the 'Rarita-Schwinger' constraints (RSC) $\gamma^t_{\alpha\beta}|t \beta \rangle_A=0 $ and the intertwining relations (cp. Appendix A)
\begin{equation}
2\theta_{\alpha A}|st \rangle_A = \gamma^s_{\alpha\beta}|t \beta \rangle_A +  \gamma^t_{\alpha\beta}|s \beta \rangle_A , \label{int11}
\end{equation}
\begin{equation}
\theta_{\alpha A}|stu \rangle_A  = \frac{i}{\sqrt{2}}\left(
\gamma^{st}_{\alpha \beta}|u\beta \rangle_A + \gamma^{us}_{\alpha
\beta}|t\beta \rangle_A +\gamma^{tu}_{\alpha \beta}|s\beta \rangle_A
\right), \label{int21}
\end{equation}
it is straightforward to calculate the action of the $SU(2)$
generators  \\ $J_A:=\frac{1}{2}\epsilon_{ABC}\theta_{\alpha
B}\theta_{\alpha C}$ on the above $Spin(9)$ singlets;
e.g.\footnote{It may be amusing to speculate about the occurrence of
the relatively large prime 13, which played a prominent role in
(bosonic) string theory.}
\[
\theta_{\alpha 1} \theta_{\alpha 2} \mathop{|||}_{44} 1 \rangle  = \frac{13}{4} |s\beta \rangle_1 |t \beta \rangle_2 |st \rangle_3
\]
and (s.b.)
\begin{equation}
\theta_{\alpha 1}\theta_{\alpha 2} \mathop{|||}_{844} 1\rangle  = -9|s\epsilon \rangle_1 |t\epsilon \rangle_2 |st \rangle_3.   \label{-9}
\end{equation}
It follows  that
\begin{equation}
\phi := \mathop{|||}_{44} 1 \rangle
+\frac{13}{36}\mathop{|||}_{844} 1 \rangle \label{state}
\end{equation}
is $Spin(9)\times SU(2)$ invariant.

Eq. (\ref{-9}) easily follows when splitting the calculation into two parts:
\[
-2\theta_{\alpha 1}\theta_{\alpha 2}|suv \rangle_1 |tuv \rangle_2 |st \rangle_3  =
\]
\[
(\gamma^{su}_{\alpha \beta}|v \beta \rangle_1  + \gamma^{uv}_{\alpha
\beta}|s \beta \rangle_1 +\gamma^{vs}_{\alpha \beta}|u \beta
\rangle_1     )   (\gamma^{tu}_{\alpha \epsilon}|v \epsilon
\rangle_2  +  \gamma^{uv}_{\alpha \epsilon}|t \epsilon \rangle_2
+\gamma^{vt}_{\alpha \epsilon}|u \epsilon \rangle_2  ) |st
\rangle_3;
\]
using the RSC and the (anti-)commutation relations between
$\gamma$'s (in particular $\gamma^{pq}=\gamma^p \gamma^q -
\delta^{pq}\bold{1}$ and $[\gamma^{pq},\gamma^r]=2\gamma^p
\delta^{qr}-2\gamma^q \delta^{pr}$), each of the nine terms becomes
proportional to $|s\epsilon \rangle_1 |t\epsilon \rangle_2
|st \rangle_3$, the respective coefficients being: $0$ for
$\gamma^{su}\gamma^{tu}$,  $\gamma^{vs}\gamma^{vt}$; $72$ for
$\gamma^{uv}\gamma^{uv}$; $3$ for  $\gamma^{su}\gamma^{vt}$,
$\gamma^{vs}\gamma^{tu}$; $-15$ for each of the remaining four
$\gamma^{su}\gamma^{uv}$, $\gamma^{uv}\gamma^{tu}$,
$\gamma^{uv}\gamma^{vt}$, $\gamma^{vs}\gamma^{uv}$.

Concerning the second part in (\ref{-9}), one notes that
\[
-\sqrt{8}i \theta_{\alpha 1} \theta_{\alpha 2}(|tuv \rangle_1|st \rangle_2 +  |st \rangle_1|tuv \rangle_2 )|suv \rangle_3 =
\]
\[
(\gamma^{tu}_{\alpha\beta}|v\beta \rangle_1 +
\gamma^{uv}_{\alpha\beta}|t\beta \rangle_1 +
\gamma^{vt}_{\alpha\beta}|u\beta \rangle_1  )
(\gamma^s_{\alpha\epsilon}|t\epsilon \rangle_2 +
\gamma^t_{\alpha\epsilon}|s\epsilon \rangle_2 )|suv \rangle_3 
\]
\[
+(\gamma^s_{\alpha\beta}|t\beta \rangle_1 +
\gamma^t_{\alpha\beta}|s\beta \rangle_1 )
(\gamma^{tu}_{\alpha\epsilon}|v\epsilon \rangle_2 +
\gamma^{uv}_{\alpha\epsilon}|t\epsilon \rangle_2 +
\gamma^{vt}_{\alpha\epsilon}|u\epsilon \rangle_2  ) |suv \rangle_3,
\]
which gives rise to 12 terms, 6 of which cancel in pairs
($\gamma^{uv}\gamma^s$, $\gamma^{tu}\gamma^t$ and
$\gamma^{vt}\gamma^t$), due to antisymmetry of $\gamma^{suv}$ (as a matrix)
and $|suv \rangle$, while (using again $\gamma^{pq}=\gamma^p\gamma^q
- \delta^{pq}\bold{1}$ and the transformation of the $\gamma^{w}$'s
as a vector when commuting with $\gamma^{pq}$) the 3 remaining terms
containing $|t\epsilon \rangle_2$, resulting in
\[
5(\gamma^v_{\beta \epsilon}| s \beta \rangle_1 | u \epsilon \rangle_2  -  \gamma^u_{\beta \epsilon}| s \beta \rangle_1 | v \epsilon \rangle_2 )|suv \rangle_3,
\]
are cancelled by those arising from the terms containing $|t\beta
\rangle_1$.

Perhaps it is worth noting that the image (under the action of any
of the $SU(2)$ generators, say $J_3$) of $\mathop{|||}\limits_{84} 1\rangle$ (which
is \emph{not} needed for the $Spin(9)\times SU(2)$ singlet) is,
using $\gamma^{stpq}\propto \gamma^{xyzvw}\epsilon_{xyzvwstpq}$,
proportional to \\ $\gamma^s_{\beta \epsilon}|u\beta \rangle |v\epsilon
\rangle |suv \rangle $ (and that the apparent 'puzzle' of the $Spin(9)$
singlet \\ $\gamma^{suv}_{\beta \epsilon}|t\beta \rangle |t\epsilon
\rangle |suv \rangle $ \emph{not} entering these considerations
is 'resolved' by observing that only the cyclically invariant
combination $\mathop{|||}\limits_{844}1\rangle$ was considered - in which
$\gamma^{suv}_{\beta \epsilon}$ terms appear \emph{symmetrized} in
$\beta$ and $\epsilon$, i.e. cancelling each other).

\section{The construction of $\phi$ out of $Spin(7)\times SU(2)$  singlets }
Here we reproduce the result obtained in the previous section
using an independent approach. Our strategy is to take advantage of the natural
$Spin(7)$ covariance of fermionic creation operators
$\lambda_{\alpha A}$, $\alpha=1,\ldots,8$, corresponding to the full $Spin(9)\times SU(2)$
model. To do so we first write the fermionic $Spin(9)$ generators
$M_{st}=\frac{1}{4}\gamma^{st}_{\hat{\alpha} \hat{\beta}}\theta_{\hat{\alpha}
A}\theta_{\hat{\beta} A}$ as ($i,j=1,\ldots,7$)
\[
M_{ij}=\frac{1}{2}\Gamma^{ij}_{\alpha \beta}\lambda_{\alpha A}\lambda^{\dagger}_{\beta A}, \ \ \ \ M_{j8}=\frac{i}{4}\Gamma^j_{\alpha \beta}(\lambda_{\alpha A}\lambda_{\beta A}+
\lambda^{\dagger}_{\alpha A}\lambda^{\dagger}_{\beta A} ),
\]
\[
M_{89}= -\frac{i}{2}\left(\lambda_{\alpha A}\lambda^{\dagger}_{\alpha A}-12 \right) \ \
\ \ M_{j9}=-\frac{1}{4}\Gamma^j_{\alpha \beta}(\lambda_{\alpha A}\lambda_{\beta
A}-\lambda^{\dagger}_{\alpha A}\lambda^{\dagger}_{\beta A} ),
\]
where we use the conventions for $\gamma^s$ as given in Appendix A.

The condition $M_{ij}\phi=0$ is the $Spin(7)$ invariance of $\phi$ while  $M_{89}\phi=0$
tells us that $\phi\in \mathcal{F}_{12}$ (where $\mathcal{F}_{n_F}$ denotes the sector with $n_F$ fermions). Therefore we
are led to search for a combination of $Spin(7) \times SU(2)$
invariant states in $\mathcal{F}_{12}$  such that
\begin{equation}
\Gamma^i_{\alpha \beta}\lambda_{\alpha A}\lambda_{\beta A}\phi = \Gamma^i_{\alpha \beta}\lambda^{\dagger}_{\alpha A}\lambda^{\dagger}_{\beta A} \phi=0. \label{spin7}
\end{equation}

\subsection{The number of $Spin(7)\times SU(2)$ invariant states}

In order to solve Eqn. (\ref{spin7}) we attempt to list all $Spin(7)
\times SU(2)$ invariant states in $\mathcal{F}_{12}$. In doing so it
is helpful to first calculate the number $D_{n_F}$ of such states
appearing in $\mathcal{F}_{n_F}$ for arbitrary $n_F$. This is done
following the lines of \cite{MT} by writing $\mathcal{F}_{n_F}$ as
\[
\mathcal{F}_{n_F} =  Alt(\otimes_{l=1}^{n_F}F_l^{s=1/2})=\mathcal{F}_{n_F, \thinspace j=0} \oplus \mathcal{F}_{n_F, \thinspace j=1/2} \oplus \ldots ,
\]
where $F_l^{s=1/2}$ is a vector space spanned by $\lambda_l^{s=1/2}| 0 \rangle$ (operators $\lambda^{s=1/2}_l$ are
assumed to carry spin $s = 1/2$ of $SO(7)$) and where $\mathcal{F}_{n_F, \thinspace j}$ is $\mathcal{F}_{n_F}$ projected
into subspaces with given $SO(7)$ angular momentum. Therefore the dimensions of subspaces with angular momentum $j = 0$ are

\[
D^{Spin(7)\times SU(2)}_{n_F}= \int d \mu_{SO(7)} \underbrace{\chi^{SO(7) \ j=0}}_{=1}\int d \mu_{SU(2)} \chi_{Alt}^{[n_F]}(R),
\]
where $d \mu_{SO(7)}$ and $d \mu_{SU(2)}$ are $SO(7)$ and $SU(2)$
invariant measures, $R$ is the adjoint representation of $SU(2)$ and
$j=1/2$ representation of $SO(7)$, i.e. $R=R^{SO(7),\thinspace j=1/2}\otimes
R^{SU(2), \thinspace j=1}$. The characters $\chi$ can be read off directly from
the Weyl character formula  while the
antisymmetric power of $\chi(R)$ is given by the Frobenius formula (see e.g \cite{BBB})
\[
\chi_{Alt}^{[n_F]}(R)=\sum_{\sum_k k i_k =n_F}(-1)^{\sum_k i_k} \prod_{k=1}^{n_B} \frac{1}{i_k!}\frac{\chi^{i_k}(R^k)}{k^{i_k}},
\]
(here $R$ is considered as a matrix). Taking all into consideration we find that the generating function for the numbers $D_{n_F}$ is
\[
\sum_{n_F=0}^{24}D_{n_F}b^{n_F}=1 + 2b^4 + 5b^8 + 7b^{12} + 5b^{16} + 2b^{20} + b^{24}.
\]
Note the duality between $\mathcal{F}_{n_F}$ and
$\mathcal{F}_{24-n_F}$, i.e. the particle-hole symmetry.
\subsection{The construction of $Spin(7)\times SU(2)$ invariant states and $\phi$}

We now proceed to construct the $Spin(7)\times SU(2)$ invariant
states in $\mathcal{F}_4$, $\mathcal{F}_8$ and finally in
$\mathcal{F}_{12}$ (note that there are no such states in
$\mathcal{F}_{4n+2}$, $n=0,\ldots,5$ and $\mathcal{F}_{2n+1}$,
$n=0,\ldots,11$). Let us first consider operators
\[
b_{AB}:=\lambda_{\alpha A}\lambda_{\alpha B}, \ \ \ \ b^i_{AB}:=\Gamma^i_{\alpha\beta}\lambda_{\alpha A}\lambda_{\beta B}.
\]
In the $\mathcal{F}_4$ sector there are only two such states for which we choose
\[
v_1|0\rangle, \ \ \ \ v_2|0\rangle \ \ \ \
v_1:=b_{AB}b_{AB} \ \ \ \ v_2:=b^i_{AB}b^i_{AB}.
\]
They are not orthogonal as the overlap matrix $G=[\langle 0|v_i^{\dagger}v_j|0\rangle]_{i,j=1,2}$ is (see Appendix B)
\begin{equation}
G=\left( \begin{array}{cc}
                        1344 & 2688      \\
                        2688  & 45696 \\
                              \end{array} \right), \ \ \ \ \det G \ne 0.  \label{G}
\end{equation}
We developed a symbolic programme written in Mathematica to confirm
that indeed $G$ is of this form.

In the $\mathcal{F}_8$  fermion sector there are 5 invariant states. Three of them are simply
\[
v_1v_1|0\rangle, \ \ \ \ v_1v_2|0\rangle, \ \ \ \ v_2v_2|0\rangle  .
\]
The remaining two are e.g.
\[
w_1|0\rangle, \ \ \ \ w_2|0\rangle, \ \ \ \
w_1=b^i_{AB}b^i_{BC}b^j_{CD}b^j_{DA}, \ \ \ \ w_1=b^i_{AB}b^j_{BC}b^i_{CD}b^j_{DA}.
\]
We checked in Mathematica that these states are linearly independent.

Finally in the $\mathcal{F}_{12}$ sector we should have 7 states. Considering the previous sectors we can construct already 8. They are
\[
v_1v_1v_1| 0 \rangle, \ \ \ \  v_1v_1v_2| 0 \rangle, \ \ \ \  v_1v_2v_2| 0 \rangle, \ \ \ \  v_2v_2v_2| 0 \rangle,
\]
\[
v_1w_1| 0 \rangle, \ \ \ \ v_2w_1| 0 \rangle, \ \ \ \ v_1w_2| 0 \rangle, \ \ \ \ v_2w_2| 0 \rangle.
\]
Accordingly there should be one relation between
them. Indeed, we found that
\[
4358 v_1v_1v_1| 0 \rangle + 2652 v_1v_1v_2| 0 \rangle + 984 v_1v_2v_2| 0 \rangle + 63 v_2v_2v_2| 0 \rangle - 528v_1w_1| 0 \rangle
\]
\[
-88 v_1w_2| 0 \rangle + 24 v_2w_1| 0 \rangle - 152 v_2w_2 | 0 \rangle= 0,
\]
and that there are no other identities among these 8 states. Therefore we can choose the $\mathcal{F}_{12}$ basis to be e.g.
\[
r_1=v_1v_1v_1| 0 \rangle, \ \ \ \  r_2=v_1v_1v_2| 0 \rangle, \ \ \ \  r_3=v_1v_2v_2| 0 \rangle, \ \ \ \  r_4=v_2v_2v_2| 0 \rangle,
\]
\[
r_5=v_1w_1| 0 \rangle, \ \ \ \ r_6=v_2w_1| 0 \rangle, \ \ \ \ r_7=v_1w_2| 0 \rangle.
\]
Finally we checked that there exists a unique combination of $r_i$ such that Eqn. (\ref{spin7}) is satisfied. The result reads
\[
\chi= 326304r_1 + 488136r_2+ 72612r_3 + 1377r_4 + 114576r_5 - 176528r_6 + 10296r_7,
\]
and is proportional to (\ref{state}).

\section{Outlook}

Having solved a (physically relevant) representation theoretic
question, let us now make a comment on 
the problem of determining the full zero-energy eigenfunction
$\Psi$ of the Hamiltonian
$$
    H = -\Delta + \frac{1}{2}(\epsilon_{ABC}x_{sB}x_{tC})^2
    + i x_{sC} \epsilon_{ABC} \gamma^s_{\hat{\alpha} \hat{\beta}} \theta_{\hat{\alpha} A}\theta_{\hat{\beta} B},
$$
which on the physical space of $SU(2)$ invariant states is equal to
the square of each of the supercharges
$$
    \mathcal{Q}_{\hat{\beta}} = -i\partial_{sA}\gamma^s_{\hat{\beta} \hat{\alpha}} \theta_{\hat{\alpha} A}
    + \frac{1}{2}\epsilon_{ABC}x_{sB}x_{tC}\gamma^{st}_{\hat{\beta} \hat{\alpha}} \theta_{\hat{\alpha} A}
    = D_{\hat{\beta}} + V_{\hat{\beta}}.
$$

Due to elliptic regularity (see e.g. \cite{browder}), any solution to
$H\Psi = 0$ must be smooth.
Accordingly, one can, around the origin, write $\Psi$
in terms of a power series in the coordinates,
$$
    \Psi(x) = \sum_{k=0}^{N} \Psi^{(k)}(x) = \psi^{(0)} + x_{tA} \psi^{(1)}_{tA}
    + \frac{1}{2} x_{tA} x_{uB} \psi^{(2)}_{tA,uB} + \ldots + \Psi^{(N)}(x),
$$
with $\psi^{(k)}_{t_1A_1 \ldots t_kA_k} \in \mathcal{H}$,
and $\Psi^{(k)}$ vanishing to order $k$ at $x=0$.

Examining the equations $\mathcal{Q}_{\hat{\beta}} \Psi = 0$ to each order in the coordinates, we find
\[
D_{\hat{\beta}}\Psi^{(1)} = 0, \ \ \ \  D_{\hat{\beta}}\Psi^{(2)} = 0,  \ \ \ \     D_{\hat{\beta}}\Psi^{(k+3)} + V_{\hat{\beta}} \Psi^{(k)} = 0, \ \ \ \ k=0,1,2,\ldots
\]
i.e.
\begin{equation*} \label{init_1}
    \gamma^t_{\hat{\beta} \hat{\alpha}} \theta_{\hat{\alpha} A} \psi^{(1)}_{tA} = 0,
\end{equation*}
\begin{equation*} \label{init_2}
 \gamma^t_{\hat{\beta} \hat{\alpha}} \theta_{\hat{\alpha} A} \psi^{(2)}_{tA,uB} = 0,
\end{equation*}
$$ \gamma^t_{\hat{\beta} \hat{\alpha}} \theta_{\hat{\alpha} A} \psi^{(3)}_{tA,uB,vC}
    + \frac{1}{2}\epsilon_{ABC}\gamma^{uv}_{\hat{\beta} \hat{\alpha}} \theta_{\hat{\alpha} A} \psi^{(0)} = 0,
$$
etc. for all $\hat{\beta}$.
Note the three separate towers of equations relating
$\Psi^{(k)}$ to $\Psi^{(k+3)}$ via intertwiners.

In any case, using that $\Psi$ must be $Spin(9)$ invariant \cite{Hasler-Hoppe},
one concludes that $\psi^{(0)}$
must be a scalar multiple of the state we constructed in this paper.

\vspace{20pt} 

\noindent{\bf Acknowledgments} We would like to thank V. Bach,
G. M. Graf, R. Suter and J. Wosiek for discussions, as well as the Swedish
Research Council and the Marie Curie Training Network ENIGMA
(contract MRNT-CT-2004-5652) for financial support.

\section*{Appendix A }

In this Appendix we work with a single $\mathcal{H}_{256}$.
A Fock space representation of  $\mathcal{H}_{256}$ can be obtained by
introducing fermionic creation operators $\lambda_{\alpha}$ and
annihilation operators $\frac{\partial}{\partial \lambda_{\alpha}}=\lambda_{\alpha}^{\dagger}$ via
\[
  \theta_{\alpha}=\frac{1}{\sqrt{2}}(\lambda_{\alpha}+\lambda_{\alpha}^{\dagger}),
    \ \ \ \  \theta_{\alpha+8 \thinspace}=\frac{1}{i\sqrt{2}}(\lambda_{\alpha}-\lambda_{\alpha}^{\dagger}).
\]
A basis of the Hilbert space $\mathcal{H}_{256}$ is obtained by acting with
products of the $\lambda_{\alpha}$'s on the fermion vacuum state  $\mid 0 \rangle$ defined by $\lambda_{\alpha}^{\dagger} \mid 0 \rangle=0$. The $Spin(9)$
generators $M_{st}=\frac{1}{4}\gamma^{st}_{\hat{\alpha}
\hat{\beta}}\theta_{\hat{\alpha}}\theta_{\hat{\beta}}$, $s,t=1,\ldots,9$ where
$\gamma^{st}=\frac{1}{2}[\gamma^s,\gamma^t]$ and $\gamma^s$ are $16
\times 16$, real, symmetric matrices satisfying
$\{\gamma^s,\gamma^t\}=2\delta^{st}\bold{1}_{16 \times 16}$,
then become
\[
M_{ij}=\frac{1}{2}\Gamma^{ij}_{\alpha \beta}\lambda_{\alpha}\lambda^{\dagger}_{\beta}, \ \ \ \ M_{j8}=\frac{i}{4}\Gamma^j_{\alpha \beta}(\lambda_{\alpha}\lambda_{\beta}+
\lambda^{\dagger}_{\alpha}\lambda^{\dagger}_{\beta} ),
\]
\begin{equation}
M_{89}= -\frac{i}{2}(\lambda_{\alpha }\lambda^{\dagger}_{\alpha }-4)
\ \ \ \ M_{j9}=-\frac{1}{4}\Gamma^j_{\alpha
\beta}(\lambda_{\alpha}\lambda_{\beta}-\lambda^{\dagger}_{\alpha
}\lambda^{\dagger}_{\beta } ),
\end{equation}
when choosing
$$
    \gamma^j=\left[ \begin{array}{cc}
        0 & i\Gamma^j     \\
        -i\Gamma^j & 0   \\
    \end{array} \right], \quad
    \gamma^8=\left[ \begin{array}{cc}
        0 & \textbf{1}_{8 \times 8}       \\
        \textbf{1}_{8 \times 8} & 0   \\
    \end{array} \right], \quad
    \gamma^9=\left[ \begin{array}{cc}
        \textbf{1}_{8 \times 8} & 0       \\
        0 & -\textbf{1}_{8 \times 8}   \\
    \end{array} \right],
$$
with $\Gamma^i$ being $8 \times 8$, purely imaginary, antisymmetric
matrices satisfying $\{\Gamma^i,\Gamma^j\}=2\delta^{ij}\bold{1}_{8
\times 8}$ \footnote{Furthermore one may choose $i\Gamma^j_{\alpha 8}=\delta^j_{\alpha}$,
$i\Gamma^j_{kl}=-c_{jkl}$ with totally antisymmetric octonionic
structure constants $c_{ijk}=+1$ for
$(ijk)=(123),(165),(246),(435),(147),(367),\\(257)$.}

As already mentioned, the Hilbert space $\mathcal{H}_{256}$
decomposes into three irreducible representations whose elements
will be denoted by $|st \rangle$, $|stu
\rangle$ and $|s\hat{\alpha} \rangle$ respectively. \\

\noindent $\bold{44}$ \\
\noindent An explicit presentation of the $\bold{44}$ in terms of creation operators $\lambda_{\alpha}$ was given in \cite{Hoppe1} as follows:
\[
|i\ne j \rangle = b_ib_j|0 \rangle, \ \ \ \  |jj \rangle = \left(b_j^2-\frac{1}{9}\bold{b}^2\right)|0 \rangle,
\]
\[
|j8 \rangle=\frac{1}{2}b_j\left(1-\frac{2}{9}\bold{b}^2\right)|0 \rangle , \ \ \ \ |j9 \rangle= -\frac{i}{2}b_j\left(1+\frac{2}{9}\bold{b}^2\right)|0 \rangle,
\]
\[
|88 \rangle=\frac{1}{2}\left(|0\rangle -
\frac{2}{9}\bold{b}^2|0\rangle + |8\rangle \right) , \ \ \ \ |99
\rangle= -\frac{1}{2}\left(|0\rangle +
\frac{2}{9}\bold{b}^2|0\rangle + |8\rangle \right),
\]
\begin{equation}
|89 \rangle=-\frac{i}{2}\left(|0\rangle-|8\rangle \right) ,  \label{44}
\end{equation}
where
\[
b_j:=\frac{i}{4}\Gamma^j_{\alpha
\beta}\lambda_{\alpha}\lambda_{\beta}, \ \ \ \  \bold{b}^2:=\sum_{i=1}^7b_ib_i, \ \ \ \ |8\rangle:=\lambda_1 \ldots \lambda_8 |0\rangle.
\]

While it is convenient to work with the overcomplete set of states
$|st\rangle=|ts \rangle$, satisfying $\sum_{s=1}^9 |ss\rangle=0$ and
transforming according to
\begin{equation}
M_{st}|uv\rangle=\delta_{tu}|sv\rangle-\delta_{su}|tv\rangle+\delta_{tv}|su\rangle-\delta_{sv}|tu\rangle, \label{M44}
\end{equation}
one should be aware of the fact that they are \emph{not} orthonormal; rather
\[
\langle st|s't'\rangle = \frac{1}{2}(\delta_{ss'}\delta_{tt'}+\delta_{st'}\delta_{s't})(1-\delta_{st})(1-\delta_{s't'})+ \delta_{st}\delta_{s't'}\left(\delta_{ss'}-\frac{1}{9}\right)
\]
(in accordance with  $|s \ne t \rangle \cong \frac{1}{2}(||st\rangle + ||ts \rangle) $ ,
$|tt\rangle \cong \frac{1}{2}(||tt\rangle+||tt\rangle) - \frac{1}{9}\sum_u ||uu\rangle $, where $||st\rangle$ are unconstrained tensor product-states satisfying
$\langle st ||s't' \rangle = \delta_{ss'}\delta_{tt'}$ )

(\ref{44}) follows from (\ref{M44}) when starting with the
27-dimensional traceless symmetric $U(1)$ ($ \cong M_{89}$)-invariant
$Spin(7)$ representation containing  $|i\ne j \rangle=b_ib_j
|0\rangle$:
\[
|9j \rangle = M_{9k}|jk \rangle = -i(b_k+b_k^{\dagger})b_jb_k |0\rangle =-\frac{i}{2}(b_j+2b_k^2b_j)|0\rangle, 
\]
\[
|8j \rangle = M_{8k}|jk \rangle  =\frac{1}{2}\left(b_j - 2b_k^2b_j\right)|0\rangle, 
\]
\[
M_{9j}|j9 \rangle = |99\rangle -|jj\rangle =-\left(\frac{1}{2}+b_j^2+b_j^2 b_k^2\right)|0\rangle, 
\]
\[
M_{8j}|j8 \rangle = |88\rangle -|jj\rangle =\left(\frac{1}{2}-b_j^2+b_j^2 b_k^2\right)|0\rangle, 
\]
(with $j \ne k$, no sum) implying
\[
 |88\rangle -  |99\rangle = (1+2b_j^2 b_k^2)|0 \rangle = |0\rangle +|8\rangle,
\]
\[
 7(|88\rangle +  |99\rangle) = -2(|88\rangle +|99\rangle + \bold{b}^2|0\rangle),
\]
i.e.
\[
|88\rangle +  |99\rangle = -\frac{2}{9}\bold{b}^2|0\rangle,
\]
hence
\[
|jj\rangle = b_j^2|0\rangle-\frac{1}{9}\bold{b}^2|0\rangle.
\]
Note that one may use
\[
[b_i,b_j^{\dagger}]=\frac{1}{2}M_{ij}+\delta_{ij}\left(1-\frac{1}{4}\lambda_{\alpha}\lambda^{\dagger}_{\alpha} \right), \ \ \ \
\]
\begin{equation}
[M_{ij},b_k]=\delta_{jk}b_i-\delta_{ik}b_j,    \label{com}
\end{equation}
which follows from
\[
[Tr(\lambda A \lambda),Tr(\lambda^{\dagger} B \lambda^{\dagger})]=-4Tr(\lambda BA \lambda^{\dagger})+2Tr(AB),
\]
\begin{equation}
[Tr(\lambda A \lambda^{\dagger}),Tr(\lambda^{\dagger} B \lambda^{\dagger})]=2Tr(\lambda^{\dagger} AB \lambda^{\dagger}). \label{gcom}
\end{equation}
Consistency conditions such as ($i \ne j$)
\[
|j8 \rangle = M_{8i}|ij\rangle = \frac{1}{2}M_{j8}|88\rangle = M_{j9}|98\rangle,
\]
lead to useful (Fierz-)identities ($j \ne k$)
\[
2b_jb_k^2 |0\rangle = \frac{2}{3}b_j^3 |0\rangle = \frac{2}{9}b_j\bold{b}^2 |0\rangle = b_j^{\dagger}|8\rangle.
\]

\noindent $\bold{84}$ \\
\noindent The construction of states $|stu\rangle$ transforming according to the antisymmetric representation $\bold{84}$
can be done analogously, starting with
\[
|ijk\rangle:=\sqrt{\frac{2}{9}}(b_ib_{jk}+b_kb_{ij}+b_jb_{ki})|0\rangle,
\]
where $b_{jk}:=\frac{1}{4}Tr(\lambda\Gamma^{jk}\lambda)$. In proving
\[
\langle ijk|i'j'k'\rangle=\delta_{ii'}\delta_{jj'}\delta_{kk'}, \ \ \ \ i<j<k, \ \ \ i'<j'<k',
\]
it is helpful to use
(\ref{com}) and the commutation rules
\[
[M_{ij},b_{kl}]=\delta_{jk}b_{il}  +\delta_{il}b_{jk}-\delta_{ik}b_{jl}-\delta_{jl}b_{ik},
\]
\[
[b_k^{\dagger},b_{ij}]=\frac{i}{4}Tr(\lambda \Gamma^{ij}\Gamma^k \lambda^{\dagger}), \ \ \ \ \langle 0 | b_{ij}^{\dagger}b_{kl}|0 \rangle = \delta_{ik}\delta_{jl}-\delta_{il}\delta_{jk}
\]
which follow from (\ref{gcom}).

All other states of the $\bold{84}$ can be obtained by application of the $M_{st}$ operators on $|ijk\rangle$ using
\[
M_{st}|pqr\rangle=\delta_{tp}|sqr\rangle - \delta_{tq}|spr\rangle + \delta_{tr}|spq\rangle - \delta_{sp}|tqr\rangle + \delta_{sq}|tpr\rangle - \delta_{sr}|tpq\rangle ,
\]
from which it follows that for fixed $i$, $j$ and $k$ we have
\[
|ij8 \rangle=M_{8k}|ijk\rangle, \ \ \ \  |ij9 \rangle=M_{9k}|ijk\rangle,
\]
\[
|i89 \rangle=-M_{9j}|ij8\rangle.
\]
Again, using the independence of $k$ in the above formulas resp.
additional relations among various states defined via $b_i$ and
$b_{jk}$ one can show that
\[
|ij8 \rangle= \frac{1}{\sqrt{2}}\left(b_{ij}|0\rangle +
b^{\dagger}_{ij}|8\rangle \right), \ \ \ \ |ij9 \rangle=
-\frac{i}{\sqrt{2}} \left(b_{ij}|0\rangle -
b^{\dagger}_{ij}|8\rangle \right),
\]

\[
|i89 \rangle = -M_{9j}M_{8k}|ijk \rangle = i(b_j+b_j^{\dagger})(b_k-b_k^{\dagger})|ijk \rangle=
\]
\[
-i(b_jb_k^{\dagger}-b_jb_k +b_j^{\dagger}b_k^{\dagger}-b_j^{\dagger}b_k )|ijk \rangle;
\]
out of the four terms the second and third one each gives zero (due
to being independent of $j\ne k$, while the sum over $j$ and $k$
vanishes), while the fourth
($=i(-b_kb^{\dagger}_j-\frac{1}{2}M_{jk})|ijk\rangle=-ib_kb^{\dagger}_j|ijk\rangle$)
equals the first (with $j$ and $k$ interchanged!) - which is easily
calculated to give
$\sqrt{\frac{2}{9}}\frac{3}{2}ib_jb_{ij}|0\rangle$; so
\[
|i89 \rangle =  \frac{i}{\sqrt{2}} \left(    b_jb_{ij} + b_kb_{ik} \right) |0\rangle  =  \frac{i}{3\sqrt{2}}\sum_{l=1}^7 b_lb_{il}|0\rangle \ \ \ \ i \ne j \ne k \ne i,
\]
from which it follows that $\Gamma^{ij}_{[\alpha\beta}\Gamma^j_{\rho
\epsilon ]} +\Gamma^{ik}_{[\alpha\beta}\Gamma^k_{\rho \epsilon ]}$
(no sum, $i\ne j \ne k \ne i$) must be independent of $j$ and $k$
(true even if $j=k$).
 \\

\noindent $\bold{128}$ 

\noindent The $\bold{128}$ representation comprises all odd fermion states in $\mathcal{H}_{256}$.  As a convenient definition one may take
\[
|t\hat{\alpha} \rangle := \frac{2}{11} \gamma^s_{\hat{\alpha} \hat{\beta}}\theta_{\hat{\beta}}|st \rangle,
\]
which does transform according to
\[
M_{uv}|t\hat{\alpha} \rangle = \delta_{vt}|u\hat{\alpha} \rangle - \delta_{ut}|v\hat{\alpha} \rangle - \frac{1}{2}\gamma^{uv}_{\hat{\alpha}\hat{\beta}}|t \hat{\beta} \rangle
\]
and explicitly exhibits the crucial RSC $\gamma^t_{\hat{\alpha}
\hat{\beta}}|t\hat{\alpha} \rangle=0.$  The intertwining relation
\begin{equation}
2\theta_{\hat{\alpha}}|st \rangle = \gamma^s_{\hat{\alpha}\hat{\beta}}|t \hat{\beta} \rangle +  \gamma^t_{\hat{\alpha}\hat{\beta}}|s \hat{\beta} \rangle \label{int1}
\end{equation}
follows when using
\[
\gamma^{su}_{\hat{\alpha} \hat{\beta}}\theta_{\hat{\beta}} |tu \rangle + \gamma^{tu}_{\hat{\alpha} \hat{\beta}}\theta_{\hat{\beta}} |su \rangle = 9 \theta_{\hat{\alpha}} |st\rangle;
\]
it is also true that
\begin{equation}
\theta_{\hat{\alpha}}|stu \rangle = \frac{i}{\sqrt{2}}\left( \gamma^{st}_{\hat{\alpha}
\hat{\beta}}|u\hat{\beta} \rangle + \gamma^{us}_{\hat{\alpha} \hat{\beta}}|t\hat{\beta} \rangle
+\gamma^{tu}_{\hat{\alpha} \hat{\beta}}|s\hat{\beta} \rangle \right), \label{int2}
\end{equation}
respectively
\[
|t\hat{\alpha} \rangle = \frac{i\sqrt{2}}{42}\gamma^{sv}_{\hat{\alpha} \hat{\beta}}\theta_{\hat{\beta}}|svt \rangle,
\]
-which of course could have alternatively been used to define $|t\hat{\alpha} \rangle$.\\

\noindent $\bold{Intertwiners}$ \\
The above intertwining relations (\ref{int1}) and (\ref{int2}) as well as the ones below (explicitly checked on the computer), we believe to be crucial for the construction of the full zero energy state;
\[
\gamma^{stu}_{\hat{\alpha} \hat{\beta}} \theta_{\hat{\beta}} |stu \rangle =0,
\]
\[
|t\hat{\alpha} \rangle=\gamma^t_{\hat{\alpha}\hat{\beta}}\theta_{\hat{\beta}}|tt \rangle,
\]
\[
|stu \rangle =\frac{i}{44\sqrt{2}} \left( \theta \gamma^{stv}\theta |uv\rangle + \theta \gamma^{tuv}\theta |sv\rangle + \theta \gamma^{tuv}\theta |sv\rangle   \right),
\]
\[
|st \rangle =\frac{i}{168\sqrt{2}} \left( \theta \gamma^{suv}\theta |tuv\rangle + \theta \gamma^{tuv}\theta |suv\rangle   \right),
\]
\[
\gamma^{su}_{\hat{\alpha} \hat{\beta}}\theta_{\hat{\beta}} |tu
\rangle - \gamma^{tu}_{\hat{\alpha} \hat{\beta}}\theta_{\hat{\beta}}
|su \rangle = \frac{11}{6 \sqrt{2} i} \gamma^u_{\hat{\alpha}\hat{\beta}} \theta_{\hat{\beta}} |stu\rangle,
\]
\[
\gamma^{sv}_{\hat{\alpha} \hat{\beta}}\theta_{\hat{\beta}}|tuv \rangle +
\gamma^{uv}_{\hat{\alpha} \hat{\beta}}\theta_{\hat{\beta}}|stv \rangle +
\gamma^{tv}_{\hat{\alpha} \hat{\beta}}\theta_{\hat{\beta}}|usv \rangle =
9\theta_{\hat{\alpha}}|stu \rangle,
\]
\[
\gamma^s_{\hat{\alpha}\hat{\beta}} \theta_{\hat{\beta}} |su \rangle = \frac{11i\sqrt{2}}{84}\gamma^{st}_{\hat{\alpha}\hat{\beta}}\theta_{\hat{\beta}} |stu \rangle.
\]

\section*{Appendix B}

In this Appendix we derive the form of the Gram matrix (\ref{G}) related to the $\mathcal{F}_4$ sector. A similar procedure can be applied for the other sectors.

 We
will use the following notation
\[
(\lambda_A A \lambda_B):=\lambda_{A\alpha}A^{\alpha \beta}\lambda_{B\beta}, \ \ \ \ (\lambda_A A \lambda^{\dagger}_B):=\lambda_{A\alpha}A^{\alpha \beta}\lambda^{\dagger}_{B\beta},
\]
\[
b_{AB}:=(\lambda_A \bold{1} \lambda_B), \ \ \ \
b^i_{AB}:=(\lambda_A \Gamma^i \lambda_B), \ \ \ \
b^{ij}_{AB}:=(\lambda_A \Gamma^{ij} \lambda_B),
\]
\[
M_{AB}:=(\lambda_A \bold{1} \lambda^{\dagger}_B), \ \ \ \ M^i_{AB}:=(\lambda_A \Gamma^i \lambda^{\dagger}_B), \ \ \ \ M^{ij}_{AB}:=(\lambda_A \Gamma^{ij} \lambda^{\dagger}_B).
\]
It is now useful to write down the generalization of commutation
relations (\ref{gcom}) for operators involving $\lambda_{\alpha A}$
and $\lambda^{\dagger}_{\alpha A}$ with color indices. We have
\[
[(\lambda_A A \lambda_B),(\lambda^{\dagger}_C B \lambda^{\dagger}_D)]=-(\lambda_C BA \lambda^{\dagger}_B)\delta_{AD}-(\lambda_D B^TA^T \lambda^{\dagger}_A)\delta_{BC}
\]
\[
+(\lambda_D B^TA \lambda^{\dagger}_B)\delta_{AC}+
(\lambda_C BA^T \lambda^{\dagger}_A)\delta_{BD}-(AB^T)\delta_{AC}\delta_{BD}+(AB)\delta_{AD}\delta_{BC},
\]
\[
[(\lambda_A A \lambda^{\dagger}_B),(\lambda_C B \lambda_D)]=(\lambda_A AB \lambda_D)\delta_{BC}-(\lambda_A AB^T \lambda_C)\delta_{BD}.
\]
The commutators of the $b-b$ type are now
\[
[b^{\dagger}_{AB},b_{CD}]=M_{CB}\delta_{DA}+M_{DA}\delta_{CB}-M_{DB}\delta_{CA}-M_{CA}\delta_{BD}+8(\delta_{AC}\delta_{BD}-\delta_{AD}\delta_{BC}),
\]
\[
[b^{\dagger}_{AB},b^i_{CD}]=M^i_{CB}\delta_{AD}-M^i_{DA}\delta_{BC}+M^i_{DB}\delta_{AC}-M^i_{CA}\delta_{BD},
\]
\[
[b^{i \ \dagger}_{AB},b_{CD}]=-M^i_{CB}\delta_{AD}+M^i_{DA}\delta_{BC}+M^i_{DB}\delta_{AC}-M^i_{CA}\delta_{BD},
\]
\[
[b^{i \ \dagger}_{AB},b^j_{CD}]=M^{ij}_{CB}\delta_{AD}+M^{ij}_{DA}\delta_{CB}+M^{ij}_{DB}\delta_{CA}+M^{ij}_{CA}\delta_{DB}
\]
\[
-\delta^{ij}\left(M_{CB}\delta_{AD}+M_{DA}\delta_{CB}+M_{DB}\delta_{CA}+M_{CA}\delta_{DB}\right)+8\delta^{ij}(\delta_{AC}\delta_{BD}+\delta_{AD}\delta_{BC}),
\]
while the commutators of the $M-b$ type are
\[
[M_{AB},b_{CD}]=b_{AD}\delta_{BC}-b_{AC}\delta_{BD},
\]
\[
[M_{AB},b^i_{CD}]=b^i_{AD}\delta_{BC}+b^i_{AC}\delta_{BD},
\]
\[
[M^i_{AB},b_{CD}]=b^i_{AD}\delta_{BC}-b^i_{AC}\delta_{BD},
\]
\[
[M^i_{AB},b^j_{CD}]=b^{ij}_{AD}\delta_{BC}+b^{ij}_{AC}\delta_{BD}+b_{AD}\delta_{BC}+b_{AC}\delta_{BD}.
\]

Now it is straightforward to evaluate the scalar product $\langle 0
| v_1^{\dagger}  v_1 |0 \rangle$ with $v_1:=b_{AB}b_{AB}$, we have
\[
[b^{\dagger}_{AB},b_{CD}b_{CD}]=-4b_{AD}M_{DB}+4b_{BD}M_{DA}+28b_{AB},
\]
which implies
\[
[v_1^{\dagger},v_1]=8M_{AB}M_{AB}-128M_{AA}+16b_{AB}M_{BC}b^{\dagger}_{CA}+40b_{AB}b^{\dagger}_{AB}+1344,
\]
hence  $\langle 0| v_1^{\dagger} v_1 |0 \rangle=1344$.

To calculate the scalar product $\langle 0 | v^{\dagger}_2  v_1 | 0 \rangle$ with $v_2:=b^i_{AA}b^i_{BB}$ we need
\[
[b^{i \ \dagger}_{AB},b_{CD}b_{CD}]=-4b_{AC}M^i_{CB}+4b_{BC}M^i_{CA}-4b^i_{AB}+4\delta_{AB}b^i_{CC},
\]
which gives
\[
[b^{i \ \dagger}_{AA},b_{CD}b_{CD}]=8b^i_{AA}, \ \ \ \  [v_2^{\dagger},v_1]=16b^i_{AB}b^{i \ \dagger}_{AB}-224M_{AA}+2688,
\]
hence  $\langle 0 | v_2^{\dagger} v_1 |0\rangle=2688$

Finally the scalar product $\langle 0 | v^{\dagger}_2  v_2 | 0 \rangle$ is obtained with use of
\[
[b^{i \ \dagger}_{AA},b^j_{BB}b^j_{CC}]=8(b^j_{AA}M^{ij}_{BB}-b^i_{AA}M_{BB})+136b^i_{AA},
\]
hence
\[
[v_2^{\dagger},v_2]=16b^i_{AA}(M^{ij}_{BB}-\delta^{ij}M_{BB})b^{j \ \dagger}_{CC}
\]
\[
+32M^{ij}_{AA}M^{ij}_{BB}+224M_{AA}M_{BB}+1120M_{AA}+352b^i_{AA}b^{i \ \dagger}_{BB}+45696,
\]
therefore  $\langle 0 | v_2^{\dagger} v_2 |0\rangle=45696$.

\end{document}